\documentclass{article}
\usepackage[utf8]{inputenc}
\usepackage[margin=1in]{geometry}
\usepackage{graphicx}
\usepackage{hyperref}
\usepackage[numbers, square, comma, sort&compress]{natbib}
\usepackage{amsmath}
\usepackage{xcolor}
\usepackage{orcidlink}

\title{Graph convolutional network for predicting abnormal grain growth in Monte Carlo simulations of microstructural evolution}
\author{Ryan Cohn\,\orcidlink{0000-0002-7898-0059} Elizabeth A. Holm \orcidlink{0000-0003-3064-5769}}
\date{\today}

\begin{document}
\maketitle
\begin{center}
    Submitted to \emph{npj Computational Materials} on July 9, 2024.
\end{center}
\begin{abstract}
    Recent developments in graph neural networks show promise for predicting the occurrence of abnormal grain growth, which has been a particularly challenging area of research due to its apparent stochastic nature.
    In this study, we generate a large dataset of Monte Carlo simulations of abnormal grain growth. We train simple graph convolution networks to predict which initial microstructures will exhibit abnormal grain growth, and compare the results to a standard computer vision approach for the same task.  The graph neural network outperformed the computer vision method and achieved 73\% prediction accuracy and fewer false positives. It also provided some physical insight into feature importance and the relevant length scale required to maximize predictive performance.  Analysis of the uncertainty in the Monte Carlo simulations provides additional insights for ongoing work in this area. 
        
\end{abstract}
\section{Introduction}
  
Abnormal grain growth (AGG) occurs when the growth rate of a small subset of grains significantly exceeds that of their neighbors. The bimodal grain size distribution resulting from AGG often significantly alters the properties of the material, making it an important phenomenon for many material systems. However, due to its stochastic nature, determining the underlying causes of AGG and predicting its occurrence is notoriously difficult. Researchers have been actively investigating AGG \cite{Rollett1997, Holm2015, Nasserrafi2017, Yamanaka2019, Lu2020, Etter2002, Rinko2019, Furnish2018} and have proposed several probable mechanisms for AGG in various materials. Despite this, however, they have not been able to predict which individual grains will grow abnormally during processing. Accomplishing this would provide useful insights for better understanding AGG.

By definition, AGG is a rare event, and may occur in fewer than one out of several thousand grains in a material. Thus, observing grain growth \emph{in-situ} with enough fidelity to capture detailed information about the initial microstructure and growth trajectory of abnormal grains is experimentally infeasible. We turn to simulation to overcome this limitation.
DeCost and Holm developed Monte Carlo simulations to study AGG in an idealized material system \cite{DeCost2017}. Within each simulation, a single ``candidate grain'' for AGG has a size advantage compared to the rest of the grains in the matrix. The remaining grains are divided into two classes, one of which forms high-mobility boundaries with the candidate grain. Grain boundary energy and all other interface mobilities in the system are uniform.  Despite their simplicity, these simulations were found to replicate AGG with realistic proportions of high-mobility interfaces, and also produced abnormal grains with similar morphologies to those observed in some real materials. This suggests their use as a model system for AGG in real materials.

DeCost and Holm found that varying the fraction of high mobility boundaries altered the mode of grain growth, suggesting that non-uniform grain boundary mobility can act as a persistence mechanism that enables AGG.  They characterized AGG in a large collection of simulations, but did not attempt to predict its occurrence in individual simulations. In this study, we leverage recent advances in machine learning to develop models for this purpose.
We investigate two different approaches for predicting the mode of grain growth from an initial microstructure. Our first method uses computer vision, which has seen success in a variety of applications in materials characterization \cite{Holm2020}. After generating images representing the initial state of the simulation, we use a pre-trained convolutional neural network to generate embeddings for each image. Finally, we use a simple machine learning pipeline to predict AGG from these embeddings.  Note that there are many other image classification pipelines that we could have used. Our justification for using this approach is that it is relatively simple and has been successfully applied to other materials image classification problems \cite{Cohn2021, Kitahara2018}.

Computer vision is advantageous because microstructure is often experimentally characterized through images from optical or electron microscopy. However, there is evidence that the local grain boundary network structure influences grain growth in materials \cite{Bhattacharya2021,Conry2022}. Graph-based representations of microstructure therefore offer an attractive alternative to images.
Conventional machine learning algorithms like convolutional neural networks require regularly structured inputs and therefore cannot efficiently process graphs with varying connectivity between nodes.
To overcome this limitation, researchers developed graph convolutional networks, which have been applied to a wide range of applications including predicting molecular properties \cite{Gilmer2017}, modeling physical relations between entities \cite{Cranmer2019}, and more \cite{Bollacker1f2,Park2020,Strokach2020}. 

Graph convolutional networks apply \emph{message passing} to process local neighborhoods in the graph structure.  This is the graph equivalent of a convolution layer applying visual filters to neighborhoods of pixels in an image. As such, message passing preserves the benefits of convolutional neural networks, including reducing the number of learned parameters and capturing localized features in the graph.
In the message passing framework, a graph contains a node feature matrix  \(N\) and an edge matrix \(E\). Each row of the node feature matrix contains the feature vector for the corresponding node in the graph, \(N_{u}=\textbf{h}_u^0\). Each row of the edge matrix contains the indices of the source and target node connected by the corresponding edge, \(E_{k}=[u,v]\). An optional edge feature matrix and graph feature vector may also be included during the message-passing process, but these are omitted in this description to avoid introducing extra complexities.

During each iteration \(i\) of message passing, a hidden state, denoted \(\textbf{h}^i\), is associated with every node \(u\) on the graph. The initial hidden state, \(\textbf{h}^0\), is the original node feature matrix. Then, for each next iteration of message passing, the hidden state is updated according to the following process.

\begin{equation}
    \label{agg-intro-messagepassing}
    \textbf{h}^{i+1}_u = \text{UPDATE}\left( \textbf{h}_u^i, \text{AGGREGATE}\left( \{\text{MESSAGE}\left( \textbf{h}_v^i, \forall v \in N(u) \right) \}\right) \right)
\end{equation}

In this equation, \( v \in N(u) \) denotes all nodes \(v \) that have directional edges  to node \(u\). MESSAGE and UPDATE can be any differentiable functions, including neural networks. AGGREGATE can be any differentiable, permutation-invariant set function that produces an output with a fixed size such as min, max, mean, sum, etc. 

In plain English, message passing works through the following steps: First, for a given target node, the message of each neighboring node is computed by applying a differentiable transformation to the neighboring node's features (first iteration) or hidden state (subsequent iterations). Next, the messages from all nodes in the neighborhood are aggregated into one signal of a fixed size.  Finally, the aggregated messages, along with the current hidden state of the target node, are transformed to compute the new hidden state of the target node. Repeating this process for each node in the graph constitutes one full iteration of message passing. 

This message passing process is repeated for a specified number of iterations. Each iteration increases the distance that information propagates through the graph by one neighborhood hop. For example, after two iterations of message passing, the hidden state of the node is influenced by the original features of the node, its nearest neighbors, and its next-nearest neighbors. The final hidden states of each node are used as inputs to a function that can generate node-level outputs, such as a classification label for each node, or can be aggregated into a single state used to make graph-level predictions, for example, a prediction of a molecular property.

Wu et. al. \cite{Wu2019} hypothesized that the majority of the predictive power of message passing neural networks comes from their ability to preserve graph structure, and many choices for the MESSAGE and UPDATE functions in typical graph convolution networks are unnecessarily complex. They developed the Simple Graph Convolution (SGC) architecture to test this hypothesis. 
During each iteration of message passing in the SGC architecture, the hidden states of all nodes in the neighborhood of a target node are weighted by their degree before being averaged together. This process is repeated for K iterations.  Afterward, node classification is achieved by passing the final hidden state of a given node through a logistic regression classifier. 

Thus, as the name suggests, SGC is a much simpler model than other graph convolution approaches. Despite this, it achieves predictive performance comparable to more complex models on standard graph benchmark tests like Cora \cite{sen2008collective,McCallum2000} and CiteSeer \cite{Bollacker1f2}, supporting the hypothesis of Wu et. al. Because of its strong performance and simplicity, we select SGC as the architecture for our graph-based approach for predicting AGG. We hypothesize that modeling grain structure as a graph will provide several advantages compared to image-based representations of the initial microstructure, and the SGC model will outperform our computer vision approach for predicting AGG.

\subsection{Code and data availability}
The code used to generate the dataset of SPPARKS abnormal grain growth simulations used in this study is available at the following link: \href{https://github.com/holmgroup/spparks-meso}{https://github.com/holmgroup/spparks-meso}. The code for the machine learning experiments is available at the following link: \href{https://github.com/rccohn/deepspparks}{https://github.com/rccohn/deepspparks}.

\begin{figure}[t!]
    \centering
    \includegraphics{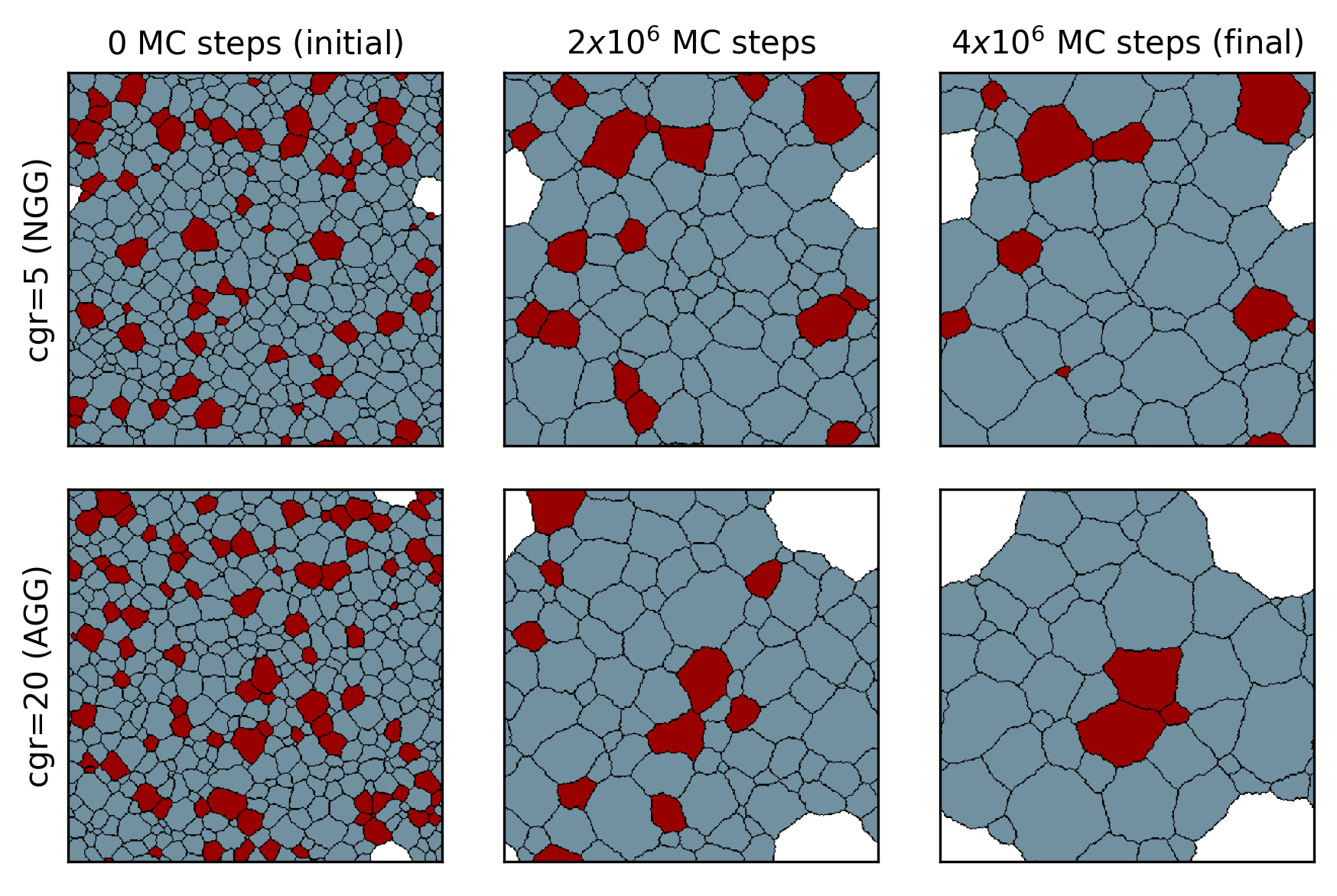}
    \caption{Evolution of sample SPPARKS grain growth simulations with CGR values of 5 (top), representing normal growth, and 20 (bottom), representing abnormal growth. Snapshots of the microstructure are shown after of 0 (left), 2,000,000 (middle), and 4,000,000 (right) iterations. The candidate grain is shown in white. In both simulations, the candidate grain ``wraps around'' the edges of the images due to the use of periodic boundary conditions.}
    \label{fig:agg:candidate-grain-example}
\end{figure}

\section{Methods}
\subsection{Dataset creation} 
\label{agg:sec:dataset_creation}
SPPARKS \cite{SPPARKS,Plimpton2009} was used to generate a dataset of ``candidate grain'' growth simulations. The simulations are summarized here, but more detailed information is available in DeCost's original study  
 \cite{DeCost2017}.

The system contains a rectangular grid of pixels with integer values. Grains consist of groups of connected pixels with the same value. One grain, called the ``candidate grain,'' has an initial size advantage compared to typical matrix grains. The rest of the matrix consists of ``red'' and ``blue'' grains. The grain boundary energy for all interfaces is uniform. The interfaces between red and candidate  grains are assigned a high mobility value, and all other interfaces are assigned the same lower mobility. During the Monte Carlo simulation, a site in the matrix is randomly selected. The probability of changing states from one grain to another, which is related to both the mobility of the interface and the energy associated with changing states, is computed. The change is accepted or rejected according to the Monte Carlo probability. This process is repeated for a very large number of iterations. During this process, the flipping of states at each pixel causes some grains to grow and other grains to shrink, modeling grain growth in real materials \cite{HOLM20032701,holm2001computer,ANDERSON1984783}. After repeating this process for the desired number of iterations, the final state of the system is recorded. 

In this study, all simulations were run with a square system with a size of 512x512 pixels using periodic boundary conditions, system energy kT = 0.9 and a duration of 4,000,000 Monte Carlo steps. 
The fraction of red grains was varied between 0.15 and 0.25. This represents the region where the occurrence of AGG is most uncertain, and about half of these trials exhibit AGG. Example grain growth trajectories from two simulations are shown in Figure \ref{fig:agg:candidate-grain-example}. The top and bottom rows show snapshots of sample trajectories for normal and abnormal grain growth, respectively.

27,588 simulations were run. The dataset was divided into a training set containing 25,588 trials, a validation set containing 1000 trials, and a test set containing 1000 trials. The ratio of the candidate grain's final size to its initial size, termed the ``candidate growth ratio'' (CGR), was used to characterize the grain growth trajectory of the system. Simulations with CGR values greater than 10 were labeled as examples of AGG, and all other simulations are labeled as normal grain growth. The threshold of 10 was selected after viewing many simulations and observing that when the CGR value exceeded 10, the growth trajectory consistently looked abnormal, with rapid bursts of growth of the candidate grain. To evaluate this subjective threshold, we also assigned labels to each sample using the criteria for AGG used in DeCost's original study \cite{DeCost2017} that is based on the expected maximum grain size in a typical grain size distribution. Both criteria assigned the same labels to 90\% of the samples in the dataset, validating our subjective judgement.

\subsection{Computer vision approach} 
Images of the initial microstructure were formatted so that the candidate grain was centered. Pixels in blue, red, and candidate grains were assigned grayscale intensity values of 1/3, 2/3 and 1, respectively. Pixels on the edges of each grain were assigned intensity values of 0 to indicate the presence of grain boundaries, and to distinguish between neighboring grains of the same type. 

\begin{figure}
    \centering
    \includegraphics[scale=0.7]{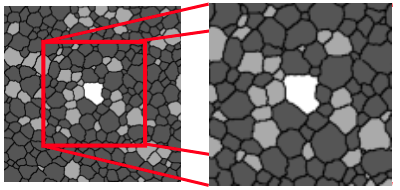}
    \caption{Grayscale image representations of the initial microstructure. Left: Resized image contains full initial state of the system. Right: Cropped image includes the local neighborhood of the candidate grain.}
    \label{fig:agg:method:imagerep}
\end{figure}

Two different approaches were used to generate images with 224x224 pixels, the required input size of the neural network used in the analysis.  In the first approach, the entire image was resized, preserving the entire initial microstructure of the simulation. To see if `zooming in' on the candidate grain results in better performance, a second set of images was evenly cropped around the center. These images typically contained about 4 complete neighborhood shells of the candidate grain.  Sample images generated after preprocessing are shown in Figure \ref{fig:agg:method:imagerep}.

Next, the VGG16 convolutional neural network \cite{Simonyan2014}, pre-trained on the ImageNet dataset \cite{Russakovsky2015}, was used to process the images. The outputs of the first fully connected layer (fc1) were  used as feature descriptors for each image. Principal component analysis (PCA) \cite{Jollife2016} was applied to reduce the dimensionality of the visual embeddings and consequently the number of learned parameters on the classifier. The number of parameters and the use of whitening were varied during training. 
Finally, a support vector machine (SVM) \cite{Cortes1995} with a radial basis kernel was trained to predict binary class labels indicating whether a given simulation would exhibit AGG. The regularization parameter C, which adjusts the trade-off between margin width and number of incorrectly classified data, was also tuned during training.

The validation set was used to perform an extensive hyperparameter search over the use of image cropping, the number of PCA components used, the application of PCA whitening, and the SVM regularization parameter C. The number of PCA components was varied between 3, which corresponds to about 25\% of the variance, and 999, preserving almost all the variance of the inputs in the small dataset. The SVM C parameter was varied between 0.01 and 100.  Results are reported for models with the set of parameters that achieved the best performance on the validation set. 

\subsection{Graph-based approach}

\begin{figure}
    \centering
    \includegraphics{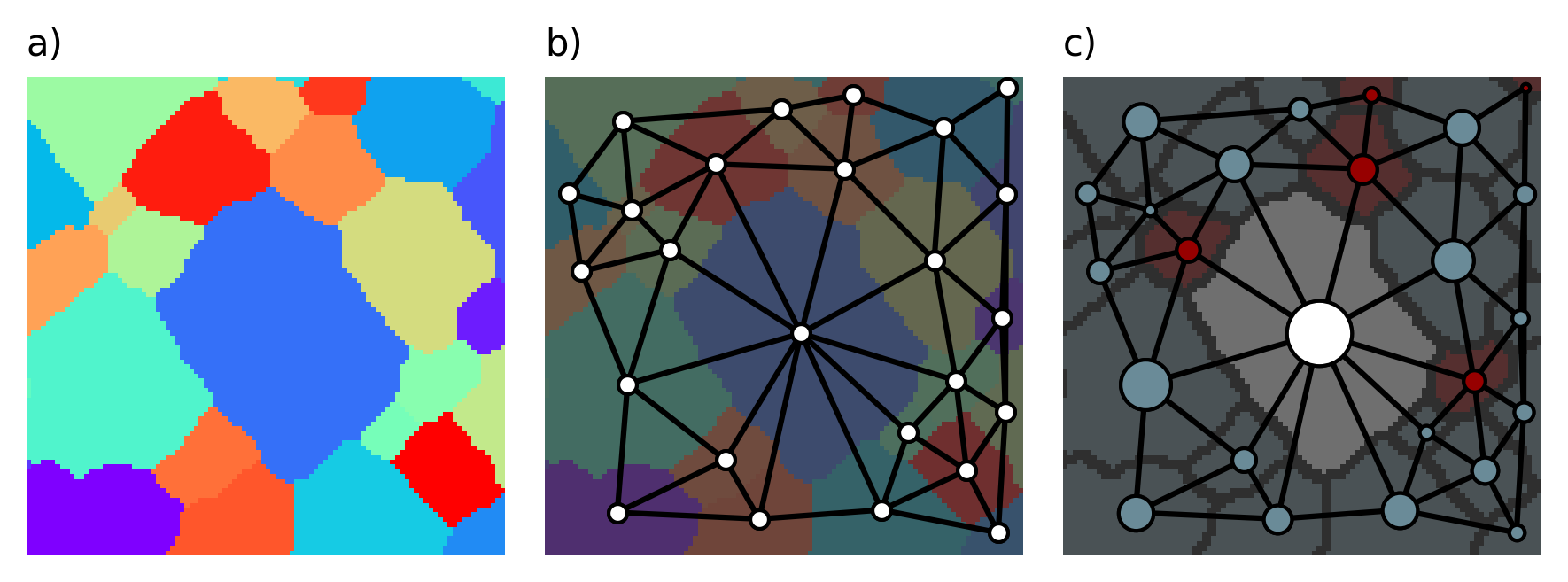}
    \caption{Schematic of graph extraction process. \textbf{a)} Output from SPPARKS. Pixel color corresponds to arbitrary grain id. \textbf{b)} Graph structure overlaid on image from a). Nodes, shown in white circles, correspond to individual grains. Edges, shown in black lines, correspond to boundaries between neighboring grains. \textbf{c)} Graph feature extraction. The color and radius of the circle for each node indicate grain ``type'' and area.}
    \label{fig:agg:methods:graph_extract}
\end{figure}
The process of extracting graph representations from images of the initial microstructure in each simulation is shown in Figure \ref{fig:agg:methods:graph_extract}. The connectivity of each node in the graph was determined from grains that overlapped after applying binary dilation to each grain.  The area, perimeter, equivalent circle diameter, major axis length, and minor axis length of the grain were measured from the pixels included in each grain. These measurements were combined with the grain's ``type'' (red/blue/candidate) and number of nearest neighbors to form the set of features for each node in the graph.

SGC models \cite{Wu2019} were trained to predict abnormal grain growth from these graph representations.
Models were trained for 2500 iterations. The performance on the training and validation sets was measured at the end of every 250 training iterations, and the model with the lowest validation loss was used for the final evaluation. During training, the negative log likelihood loss function was used to measure the distance between the predicted and true values for training examples. The Adam optimizer \cite{Kingma2015} was used to update the model parameters during backpropagation. The parameter sweep included learning rates of 0.01 and 0.005, and Adam weight decay values of \( \beta_1 = 10^{-4}\) and \( \beta_2 = 5 \times 10^{-4}\). \(K\) was varied from 1 to 4 to measure the influence of the number of neighborhood shells on the predictive performance of the model. Note that the initial system typically included about 10 neighborhood shells of the candidate grain, so differences in performances between models with different values of \( K \) should not be affected by system size.
 
\section{Results and discussion}
\subsection{Computer vision baseline}
\label{agg1:sec:results-cv}

\begin{figure}
    \centering
    \includegraphics{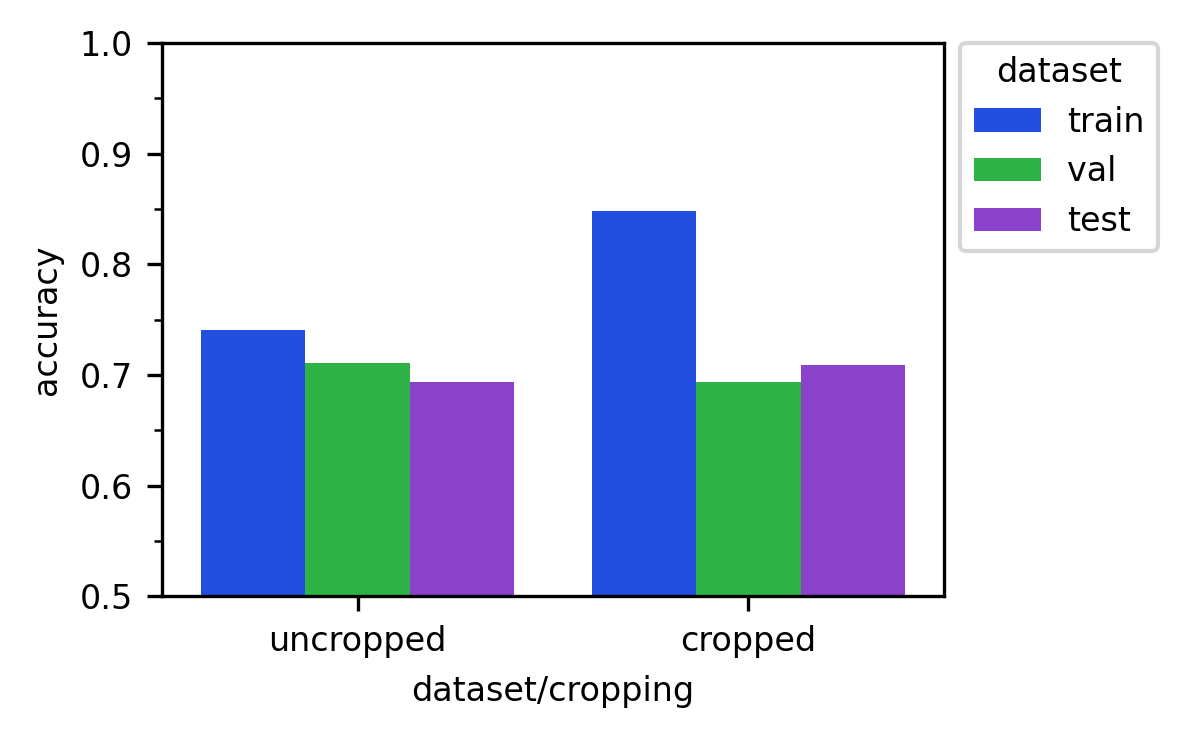}
    \caption{Training, validation, and test accuracy for SVM models trained with and without image cropping. Each result represents the model with the best validation performance after parameter tuning.}
    \label{fig:agg1-results-svm-perf}
\end{figure}

The performance of the computer vision approach is summarized in Figure \ref{fig:agg1-results-svm-perf}. All results include the use of PCA whitening, which was found to slightly but consistently increase model performance.  The SVM regularization parameter C was 0.18 for uncropped images and 0.75 for cropped images. Cropping images to only include the local neighborhood of the candidate grain increased the accuracy (defined as the fraction of model predictions that are correct) on the training set from 75\% to 85\%, but this improvement does not generalize to the validation and test sets. Regardless of whether cropping is applied, the validation or test accuracy for each model remains just under 70\%.

\begin{figure}
    \centering
    \includegraphics{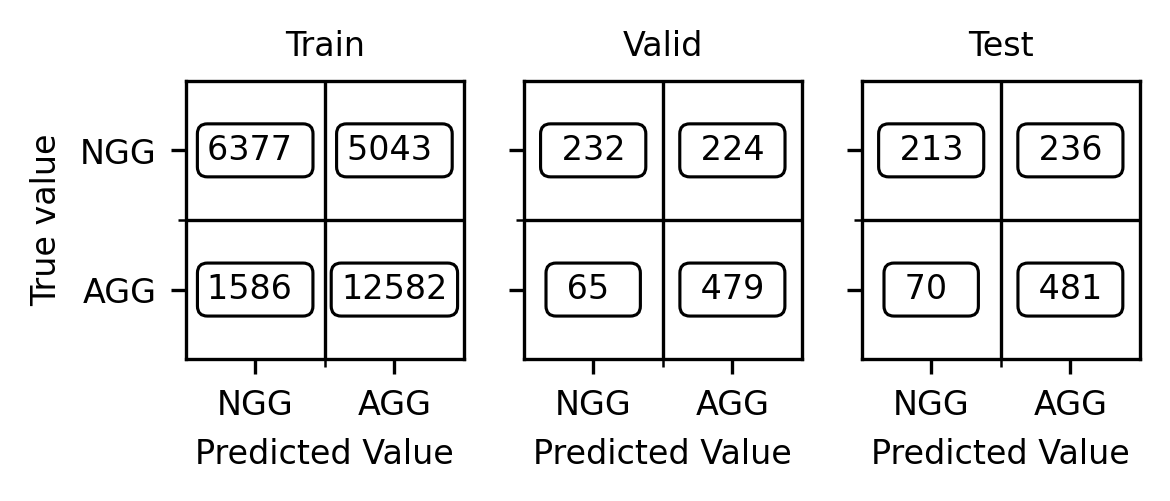}
    \caption{Confusion matrices for computer vision model trained without cropping images.}
    \label{fig:agg1-results-svm-cm}
\end{figure}

The biggest source of classification error resulted from false positive predictions of AGG in simulations that showed normal grain growth. To illustrate this point, confusion matrices for the training, validation, and test set for model predictions without the use of image cropping are shown in Figure \ref{fig:agg1-results-svm-cm}. For each of the training, validation and testing sets, false positive predictions of AGG outnumber false negative predictions by a factor of about 3, and the test precision (defined as the fraction of positive predictions of AGG that are correct) and recall (defined as the fraction of samples displaying AGG that were correctly identified by the model) were measured to be 0.671 and 0.873, respectively.

 The dataset was slightly unbalanced, with 55\% of the training set exhibiting AGG. Thus, the models appear to learn that AGG occurs more frequently than normal grain growth, which is reflected in the predictions.  To determine the impact of class imbalance on model performance, the threshold for the CGR value for AGG was adjusted to 10.945, the median value for the dataset.  Surprisingly, however, balancing the dataset did not improve the overall performance. The model achieves training, validation, and test accuracy scores of 71.7\%\%, 69.4\%, and 69.6\%, respectively. Furthermore, the biggest source of classification errors still results from false positive predictions of AGG. The relative rate of false positive predictions of AGG is about double the rate of false negative predictions, indicating that imbalanced predictions are not simply a result of seeing more examples of AGG in the training dataset.
  The test precision and recall for AGG are 0.666 and 0.781, respectively. The precision is about the same as the original predictions shown in Figure \ref{fig:agg1-results-svm-cm}. Despite having fewer false positive predictions, there are also fewer true positive predictions after balancing the dataset, and the overall rate of false positive predictions did not change. Similarly, the recall decreases after balancing the dataset due to the larger number of negative samples.

\subsection{Simple graph convolution}
\label{agg1:sec:results-sgc}

\begin{figure}
    \centering
    \includegraphics{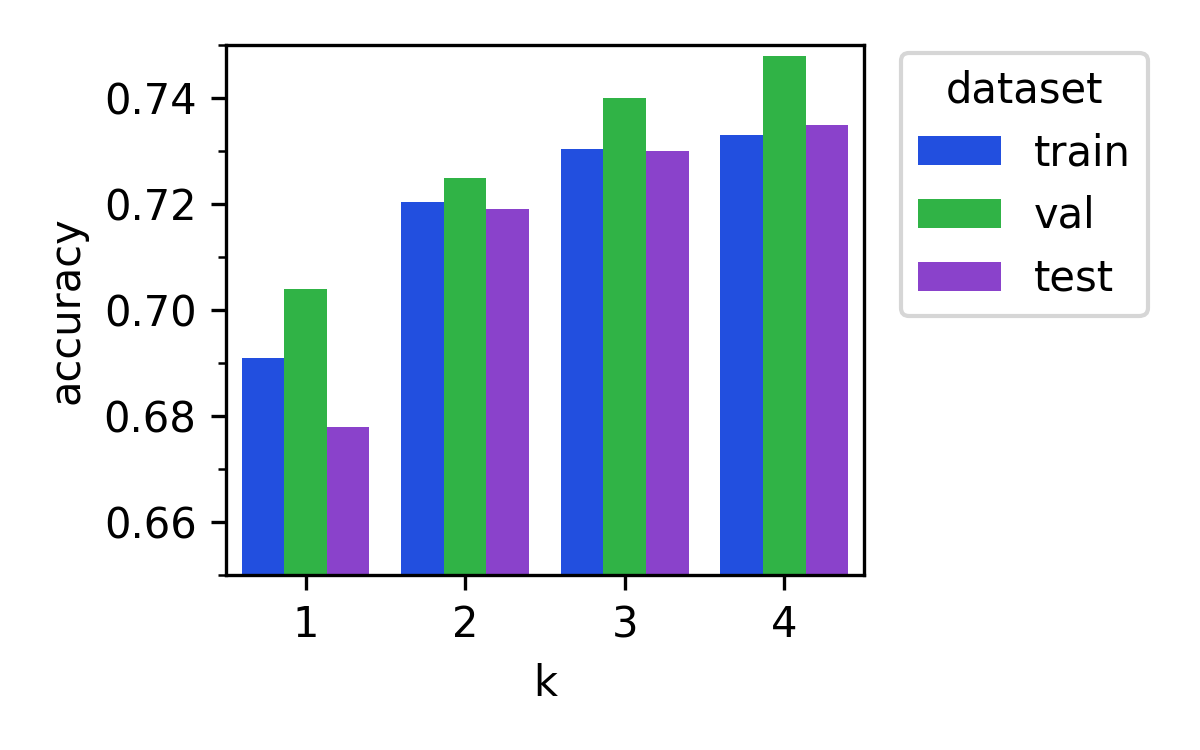}
    \caption{Accuracy scores for simple graph convolution models.}
    \label{fig:agg1-results-sgc-performance}
\end{figure}

The performance of SGC models with different values of K, the number of iterations of message passing, is shown in Figure \ref{fig:agg1-results-sgc-performance}. Model performance increases monotonically with \(K\).  The best performing model, with K=4, achieved training, validation, and testing accuracy scores of 73.3\%, 74.8\%, and 73.5\%, respectively. Thus, the graph-based approach outperforms the computer vision baseline approach. Furthermore, a SGC model trained on a subset of 1000 samples from the training set achieves comparable training, validation, and test accuracy scores.  
In other words, even with a 96\% smaller training set, the simple graph convolution model trained on still outperforms the computer vision model. This suggests that the graph representation of the microstructure preserves more relevant information than the image-based representation, despite having significantly fewer features and model parameters.

In addition to achieving a higher classification accuracy, the simple graph convolution approach demonstrates less overfitting on the training set compared to the computer vision approach. In fact, the validation accuracy is consistently about 1\% higher than the training accuracy. This is attributed to a combination of sampling error and underfitting. There may be some samples that are unusually difficult to predict that were not selected to be in the validation set, causing lower performance on the training and test sets but not the validation set. Also, with few trainable parameters, the SGC model is not able to overfit the training set, and its training performance does not exceed the performance on the validation set.  Preliminary experiments indicate that a model with more parameters performs better on the training set than the validation and test sets, supporting that the SGC underfits the data.

The performance of the model increases monotonically as K increases from 1 to 4. Increasing K=1 to K=2 gives the greatest improvement in predictive performance. As K is further increased, the improvement in classification accuracy tapers off.   Since K is equivalent to the number of neighborhood shells of the candidate grain included in the analysis, this provides some physical insight for AGG. Only considering nearest neighbors neglects to include relevant information contained in additional neighborhood shells of the candidate grain. Including additional neighborhood shells increases the performance. However, the increase in performance diminishes, and adding \(4^{\text{th}}\)-nearest neighbors only increases the accuracy of predictions by less than 1\%. Therefore, it is likely that the grains farther away from the candidate grain are too far to significantly influence its mode of growth.

\begin{figure}
    \centering
    \includegraphics{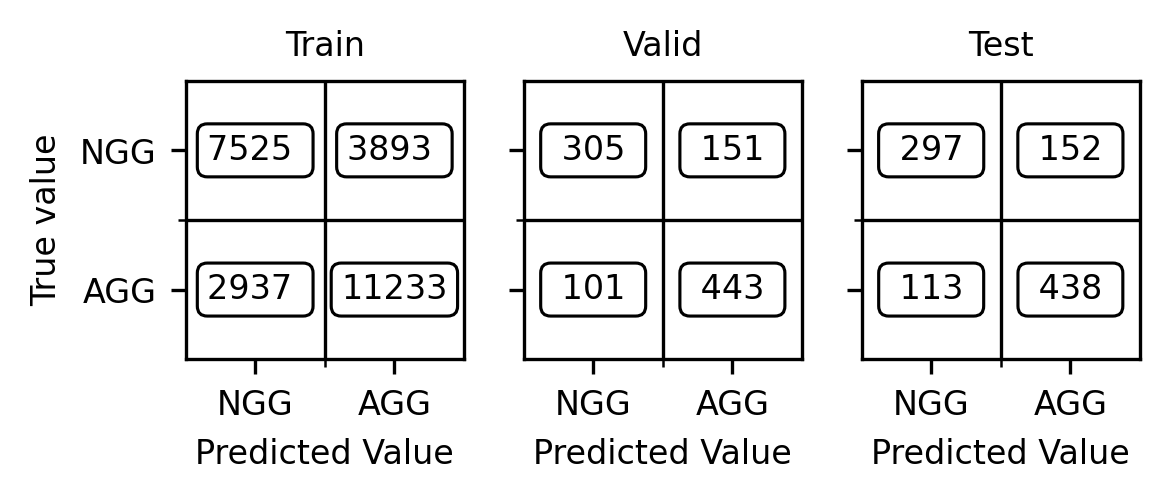}
    \caption{Confusion matrices for simple graph convolution model with K=4.}
    \label{fig:agg1-results-sgc-cm}
\end{figure}

Confusion matrices for the simple graph convolution model with K=4  on the training, validation, and test sets are shown in Figure \ref{fig:agg1-results-sgc-cm}.  Similarly to the computer vision approach, the largest source of confusion results from false positive predictions of AGG. 
However, the simple graph convolution model only shows about 30\% more false positive predictions than false negative predictions of AGG. This is significantly less than the computer vision approach, which had about three times as many false positives as false negatives. This suggests that the simple graph convolution model is less affected by the imbalanced dataset than the SVM used in the computer vision approach. The SGC model achieved both higher overall accuracy and fewer false positives, supporting our hypothesis that it outperforms the computer vision approach for predicting grain growth.

\subsection{Model interpretability: feature importance}
\label{agg1:sec-interpretability}
An ablation study was performed to probe the importance of input features for the simple graph convolution networks. Models with K=3 were trained on a subset of the dataset with 1000 training, 100 validation, and 100 test samples. The number of features was varied from 1 to 7, the total number of node features used in the study. All possible combinations of features were tested. Somewhat surprisingly, the only feature found to impact performance was grain ``type'' (red, blue, candidate), which is related to the mobility of the interfaces that a grain will form with others in the simulation. All subsets of features that included grain type yielded results similar to the full feature set, and all subsets of features that excluded grain type resulted in poor model performance.

\begin{figure}
    \centering
    \includegraphics{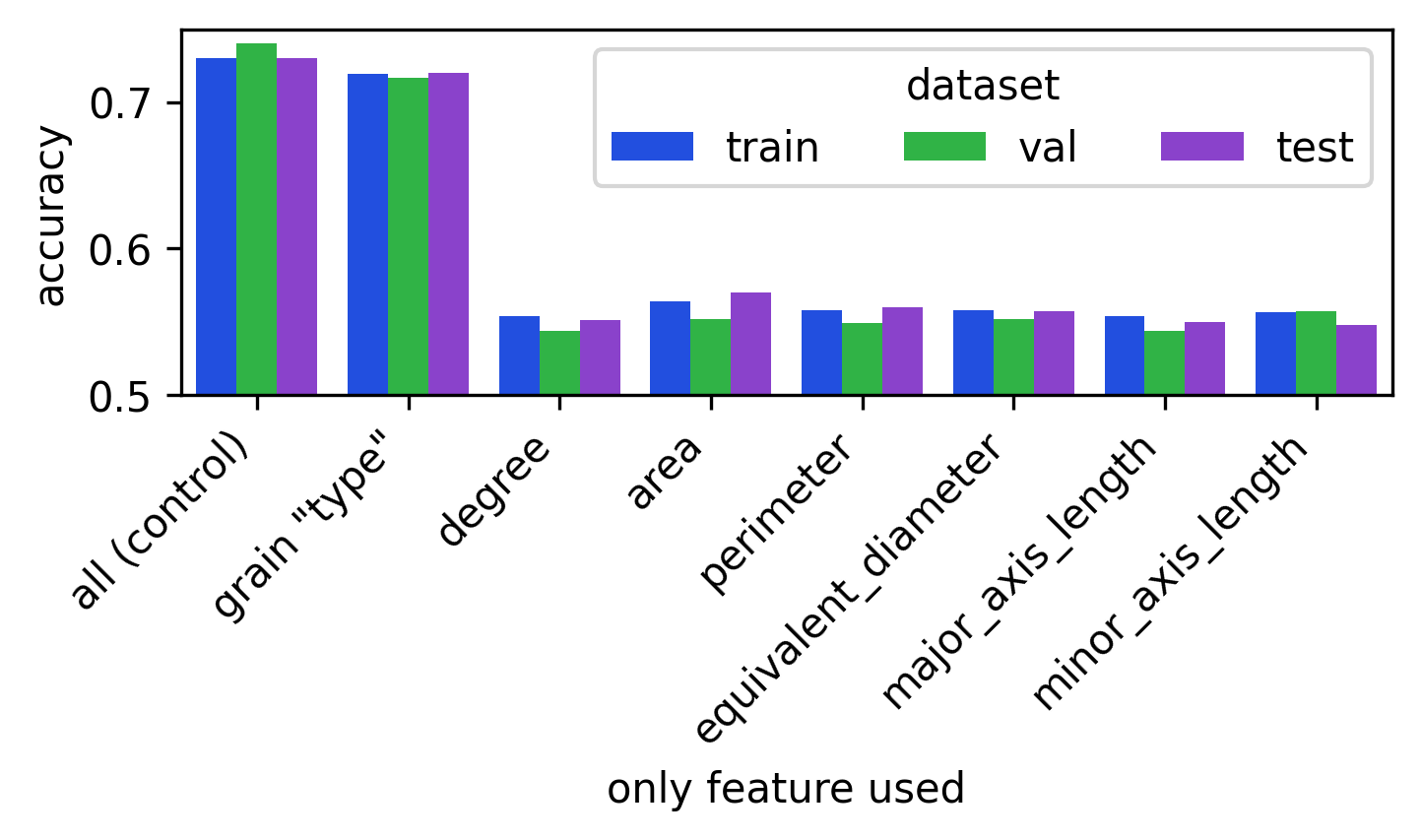}
    \caption{Performance of simple graph convolution models with K=3 trained on a single node feature}
    \label{fig:agg1:interpratability:1feat}
\end{figure}

To further investigate this finding, a series of models was trained on the full dataset using only a single feature. The results are shown in Figure \ref{fig:agg1:interpratability:1feat}.  The model trained only on grain type achieves training, validation, and testing accuracy scores of 71.9\%, 71.7\%, and 72.0\%, respectively. This is comparable to the 73\% accuracy achieved by training the model on all features. Models trained on every other feature all achieved around 55\% accuracy on the training, validation, and test subsets.  These models predict AGG in 85-100\% of samples. This indicates that they cannot extract any meaningful signal from the features, and are simply learning that the dataset contains slightly more examples of AGG than normal growth.

The feature set includes grain type, the number of nearest neighbors, and some parameters describing the size and shape of each grain. The microstructures are initialized with relatively compact and equiaxed grains with a typical grain size distribution. The candidate grain is always selected to be one of the largest initial grains. Because of this, the candidate grain's initial size, shape, and number of neighbors do not vary significantly between simulations. 
Individual matrix grains have a bigger range of size and number of neighbors, but these aren't significant enough to alter the growth trajectory of the candidate grain. In contrast, all the predictive power of the SGC model comes from grain ``type". Although there is a similar fraction of red grains in each simulation, the local distribution of grain types varies significantly. This can be seen in Figure \ref{fig:agg:candidate-grain-example}. In the example of relatively slow grain growth shown on top, most of the grains close to the candidate grain are blue, and there are only two series of several connected red grains within three neighborhood hops of the candidate grain. In contrast, the example of abnormal grain growth shown on the bottom has four series of connected grains within three neighborhood hops. During message passing, the SGC measures the local fraction of red grains near the candidate grain, allowing it to find regions where rapid bursts of growth can occur. It is somewhat surprising that the SGC does not account for both size and grain type, as combining these quantities gives a more direct measurement of the total area available for rapid bursts of growth. However, due to the geometry of the microstructure, and the fact that grain types are assigned randomly, grain size is correlated with the number of grains in a given neighborhood. As a grain's size is increased, it occupies more length along the boundary, limiting the number of other grains that can form an interface with the candidate grain.  Therefore, encoding grain size directly as a feature is redundant, and does not result in significant performance improvement of the SGC model.

It is difficult to perform a comparable feature importance assessment for the computer vision approach. However, it is interesting to note that grain type is directly encoded in the CV model via grain color. In that sense, the computer vision model has access to the most important feature as determined by the SGC model, yet it performs significantly worse than the SGC model. This suggests that grain type gains importance when combined with information about grain connectivity, which is directly encoded in the SGC model. This is consistent with our understanding of the physical mechanism of AGG, which propagates by consuming neighbor grains within the polycrystalline grain network.

\subsection{Estimation of the maximum meaningful prediction accuracy}
\label{sec:agg1:max_acc}
With a 27\% error rate, there is plenty of room for improvement in our graph-based approach for predicting AGG.  Changing the graph feature extraction or model architecture could lead to incremental improvements in performance. However, there is another factor that limits the ability to predict AGG: the data itself. 

In this study, a single growth trajectory is associated with each initial microstructure. However, the Monte Carlo Potts method is stochastic. Running the same grain growth simulation with different random seeds may result in grain growth trajectories both above and below the threshold for AGG. This uncertainty limits the ability of any model to predict a single growth trajectory for a given initial state.

\begin{figure}
    \centering
    \includegraphics{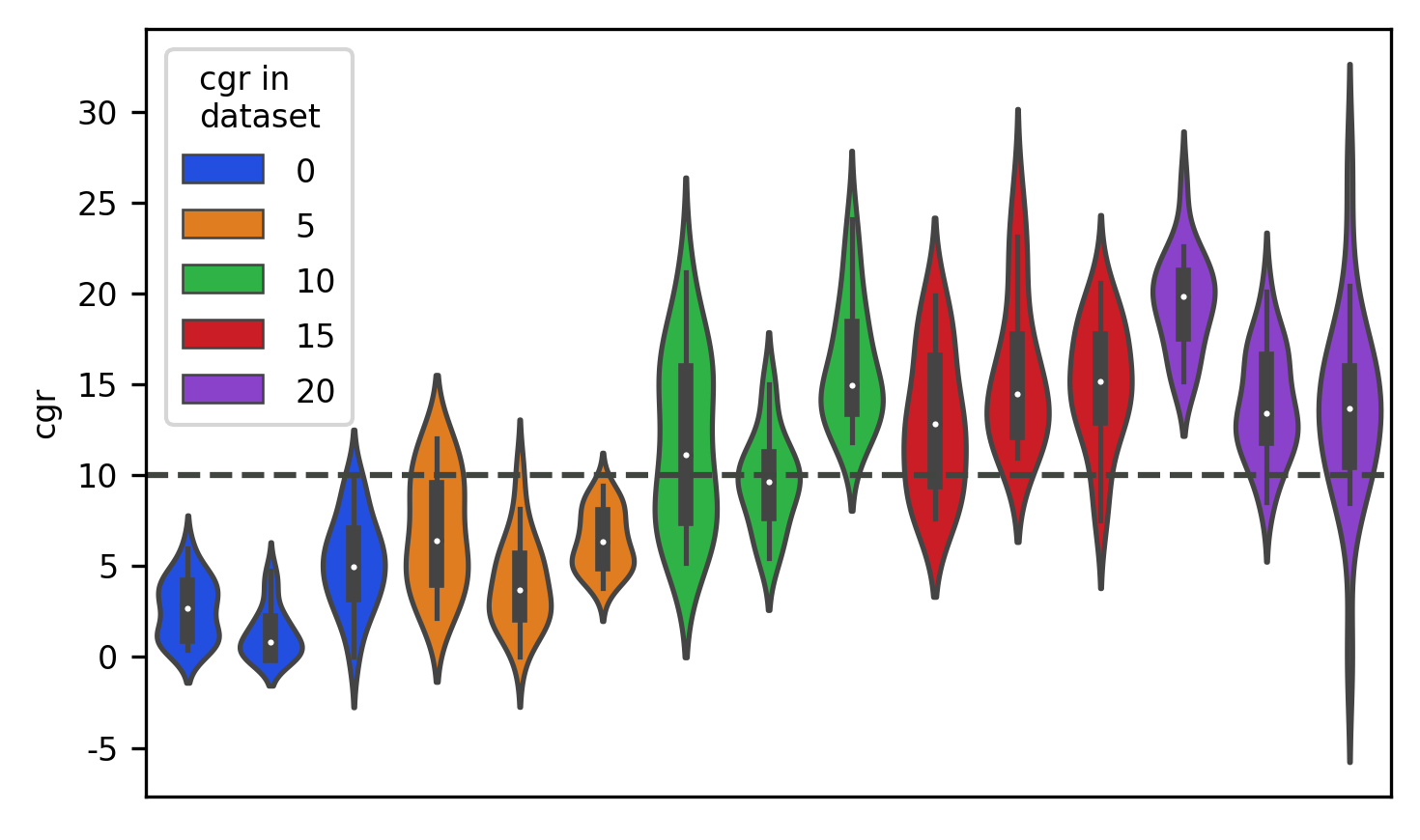}
    \caption{CGR distributions for 20 repeated simulations for each initial state.  Trials are ordered and colored by the CGR value observed in the original dataset with only one grain growth simulation. The distributions for each violin are determined from kernel density estimation and can include nonphysical negative values.}
    \label{fig:agg-results-cgrepeat}
\end{figure}

To measure the dependency of the growth trajectory of the candidate grain on the random state of the simulation, 3 trials that had reported CGR values close to 0, 5, 10, 15 and 20 were rerun with 20 different random seeds.  The distribution of CGR values for each initial state are shown in Figure \ref{fig:agg-results-cgrepeat}.  The spread of each distribution clearly demonstrates that the final size of the candidate grain is highly dependent on random state.

The first three distributions, colored blue, show the grain growth trajectories for simulations where the candidate grain was consumed by neighboring grains (CGR=0). Compared to the rest of the samples, this set of initial states has the lowest overall variation in CGR value. It is the only set of samples in which all simulations resulted in CGR values on the same side of the threshold, although one trial almost achieves AGG with a CGR value of 9.75. Despite this apparent consistency, the set of distributions contains several samples with CGR values above and below 1. This highlights an important result: For these simulations, whether the candidate grain grows or shrinks is determined entirely by random state.

The remaining sets of trials contained CGR values both above and below 10, confirming that the grain growth trajectory was at least partially determined by random state. Initial states with CGR values of 10 or 15 in the original dataset typically showed the highest amount of variation. Because they have both CGR values close to the threshold used to classify AGG and the most variation, the growth mode of these samples have the highest likelihood of changing growth modes when simulations are repeated. 

Two out of three sets of trials with original CGR values of 20, shown on the first two distributions from the right of the graph, had median CGR values below 15. This is lower than the median CGR of two sets of trials with original CGR values of 15, and one with an original CGR of 10. Both of these distributions include trials that underwent normal grain growth. From the single trial observed in the initial dataset, these samples would appear to stand out as more extreme examples of AGG. However, repeating the simulations shows that the observed grain growth trajectory was a result of a low probability growth event instead of the initial state having the conditions that enable extreme AGG.
In contrast, the remaining sample with a CGR of 20 in the initial dataset, shown in the third distribution from the right on the graph, consistently shows more extreme AGG. It has a median CGR of about 20, and none of the repeated trials demonstrated normal grain growth. Repeating grain growth simulations distinguishes this initial state as being primed for extreme AGG.

Despite the stochastic nature of Potts Monte Carlo simulations, the extent of the variation in grain growth trajectories observed from repeating simulations is surprising. Random fluctuations at grain boundaries are expected to average out after a large number of iterations to yield grain growth trajectories that are relatively consistent. Macroscopic differences in grain growth, such as the difference between the candidate grain growing by a factor of 5 or being consumed by its neighbors and eliminated, are not typically expected. 
The implication is that the Monte Carlo simulations cannot accurately model the trajectory of individual candidate grains. On the other hand, the aggregation of many simulations can provide insight about the expected trajectories, as well as the limits of predictability.

The mean value of the standard deviation of the CGR measurements for each initial state, denoted \(\sigma\), was found to be 3. A preliminary way to estimate the uncertainty in the data set and the selected criteria for AGG is to count the fraction of samples with growth ratios within \(\pm \sigma\) of the threshold used for AGG. These samples have a high probability of switching labels from AGG to normal grain growth, or vice versa, when re-run with a different random seed.

About 35\% of the samples in the training set meet this criterion. This indicates that the labels are consistent for only about 65\% of the samples in the dataset.  Assuming that the remaining 35\% have an equal chance of demonstrating normal or AGG depending on the random state, a predictive model would be expected to correctly predict the labels of half of them. Thus, for a large enough number of samples, the maximum accuracy achieved by an optimal predictive model is expected to be around 83\%. This corresponds to an error rate of 17\%, which is approximately 37\% lower than the error rate of the simple graph model developed in this study.  

\begin{figure}
    \centering
    \includegraphics{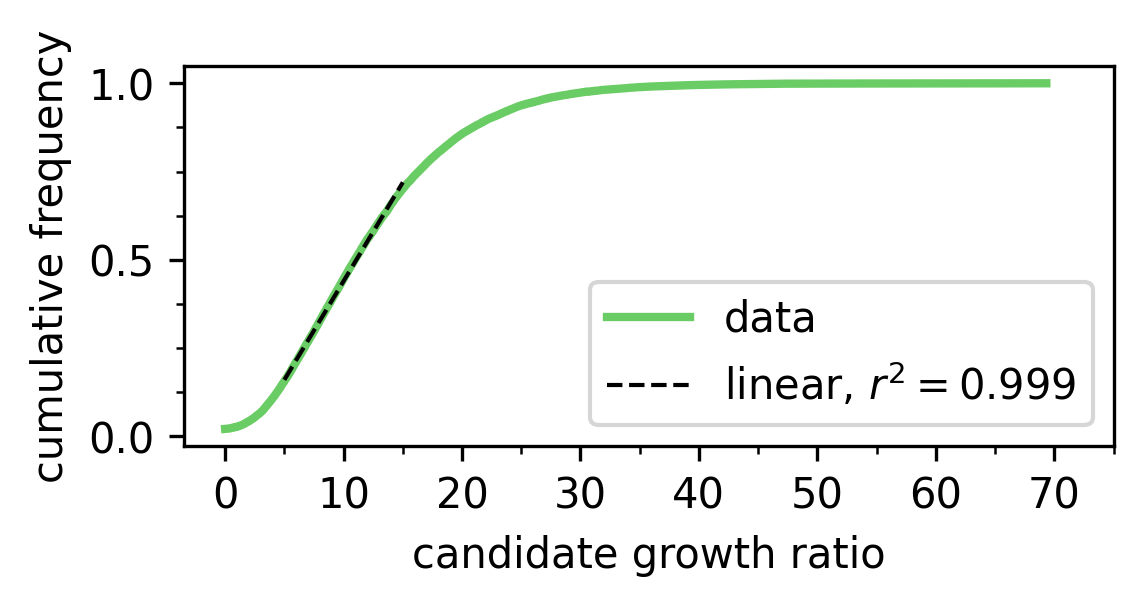}
    \caption{Cumulative distribution of CGR values in the training dataset. The linear trend for CGR values between 5 and 15 demonstrates indicates that there is no `natural' threshold for distinguishing normal and abnormal grain growth based on final grain size.}
    \label{fig:agg1:results:cgr_dist}
\end{figure}

It is possible that using a more consistent definition of AGG will lead to less uncertainty in the data. In this study, a simple threshold for the CGR, the ratio of the final area to the initial area of the candidate grain, is used to distinguish between normal and abnormal grain growth. The cumulative distribution of the CGR values in the training dataset  is shown in Figure \ref{fig:agg1:results:cgr_dist}. The distribution does not appear to be bimodal, suggesting there is no obvious threshold for grain growth mode based on grain size alone. Instead, the it is very linear near the threshold CGR value of 10. There are many samples with values close to the threshold. Therefore, the uncertainty in the labels cannot be reduced by simply moving the threshold.

This finding raises the question of whether applying a different criterion for AGG could result in more consistent labels. For example, AGG is observed to occur in sudden bursts of growth between periods of slow or no growth. Measuring these short-term bursts of growth could provide a more consistent criterion for AGG with a weaker dependence on random state. However, the extensive variation in grain growth trajectories observed in Figure \ref{fig:agg-results-cgrepeat} suggests that the occurrence of these growth events may also have a strong dependence on random state.   Determining a more consistent criterion for AGG remains an open question that requires additional research efforts.

An alternative approach that is more robust to variations caused from the random state is to repeat the grain growth simulations and characterize the resulting distribution of observed grain growth trajectories, rather than mapping the initial structure to a single outcome. This idea provides motivation for ongoing efforts to characterize and better understand AGG.

\section{Conclusions} 
SPPARKS was used to generate a large dataset of ``candidate grain" simulations of abnormal grain growth.  Manual analysis of several simulations led to a preliminary working definition of abnormal grain growth based on the relative increase in size of the candidate grain. Two approaches were applied to predict the occurrence of abnormal grain from the initial state of each simulation. The first approach utilized a computer vision and transfer learning approach. After using VGG16 pre-trained on ImageNet as a feature extractor, a kernel SVM classifier was used to predict abnormal grain growth. This approach achieved approximately 69\% accuracy on the test set. It demonstrated significant overfitting to the training set and generated a significant number of false positive predictions of abnormal grain growth. In the second approach, a simple graph convolution model with was used to predict abnormal grain growth. Despite having fewer trainable parameters, this approach achieved 73\% accuracy on the test set, corresponding to a 13\% reduction in error rate compared to the computer vision method. Additionally, this model did not show signs of overfitting and generated fewer false positive predictions of abnormal grain growth. Varying the number of iterations of message passing used by the model indicated that most of the relevant information to generate predictions is contained within the first two neighborhood shells of a given grain. Conducting a feature ablation study indicated that grain type, which is a proxy measurement of interface mobility, along with the graph structure, was sufficient to achieve maximum predictive performance.

Further analysis of the model performance led to an investigation in the intrinsic uncertainty in grain growth behavior in the Monte Carlo simulations. Repeating grain growth simulations with the same initial state and simulation parameters but with different random seeds revealed a significant variation in the final state of the system. This variation translated into significant uncertainty in the classification of grain growth trajectory as ``normal'' or ``abnormal''. Ongoing research efforts strive to investigate the use of more sophisticated feature sets and model architectures to improve predictive performance and increase the robustness of the approach to the stochastic nature of these simulations by repeating grain growth simulations with different random seeds.

\section*{Acknowledgements}
This work was supported by the U.S. Air Force Data-Driven Discovery of Optimized Multifunctional Materials Systems (D3OM2S) under cooperative agreement FA8650-19-2-5209.

\bibliographystyle{unsrtnat}
\bibliography{refs}
\end{document}